\documentclass[12pt]{article}

\usepackage{graphics}
\usepackage{amsmath}
\usepackage{epsfig}
\usepackage{cite}

\titlepage
\title{TMD parton shower effects in associated $\gamma$ + jet production at LHC}

\author{A.V.~Lipatov$^{1,\,2}$, M.A.~Malyshev$^1$, H.~Jung$^{3}$}

\begin{document}

\maketitle

\begin{center}

{\it $^1$Skobeltsyn Institute of Nuclear Physics, Lomonosov Moscow State University, 119991 Moscow, Russia}\\
{\it $^2$Joint Institute for Nuclear Research, 141980 Dubna, Moscow Region, Russia}
{\it $^3$Deutsches Elektronen-Synchrotron, 22603 Hamburg, Germany}\\

\end{center}

\vspace{0.5cm}

\begin{center}

{\bf Abstract }

\end{center} 

\indent
We investigate associated prompt photon and 
hadronic jet production at the LHC energies using the $k_T$-factorization approach.
Our consideration is based on the $\mathcal O(\alpha\alpha_s^2)$ off-shell gluon-gluon fusion
subprocess $g^*g^*\to \gamma q\bar q$ and several subleading quark-initiated contributions 
from $\mathcal O(\alpha\alpha_s)$ and $\mathcal O(\alpha\alpha_s^2)$ subprocesses, 
taken into account in the conventional (collinear) QCD factorization. 
The transverse momentum dependent (or unintegrated) gluon densities in a proton are derived from 
Catani-Ciafaloni-Fiorani-Marchesini (CCFM) evolution equation. 
We achieve reasonably good agreement with the experimental data taken 
by CMS and ATLAS Collaborations and demonstrate the importance of initial state parton 
showers for jet determination in the $k_T$-factorization approach.

\vspace{1.0cm}

\noindent
PACS number(s): 12.38.-t, 12.38.Bx, 14.70.Bh

\newpage

\section{Motivation} \indent

Investigation of prompt photon and associated hadronic jet production is an important topic of modern experimental and 
theoretical research\cite{1,2,3,4,5,6}. The photons are called prompt, if they originate from the hard partonic subprocess, 
rather than from secondary decays.
Such events provide a direct probe of the hard
subprocess dynamics since the produced photons are largely insensitive to the effects of
final-state hadronization.
The measured $\gamma$ + jet total and differential cross sections are sensitive to the quark and gluon
densities in the proton over the whole kinematical region of 
longitudinal momentum fraction $x$ and hard scale $\mu^2$
and represents an important background to many processes involving photons in the
final state, including Higgs boson production (in diphoton decay mode). Thus, it is 
essential to have accurate QCD predictions for corresponding cross sections.

The reported measurements\cite{1,2,3,4,5,6} are in agreement with the 
results of next-to-leading-order (NLO) perturbative QCD calculations performed 
using \textsc{jetphox} Monte-Carlo event generator\cite{7}.
The leading-order (LO) calculations based on the Monte-Carlo event generator \textsc{sherpa}\cite{8}, which incorporates 
higher-order tree level matrix elements and parton shower modeling,
also agree well with the measurements\cite{1,2,3,4,5,6}.
An alternative description of $\gamma$ + jet data can be achieved in the 
framework of the high-energy QCD factorization\cite{9}, or $k_T$-factorization approach\cite{10}.
This approach is based on the Balitsky-Fadin-Kuraev-Lipatov (BFKL)\cite{11} or 
Ciafaloni-Catani-Fiorani-Marchesini (CCFM)\cite{12} gluon evolution equations and has
certain technical advantages in the ease of including higher-order QCD radiative corrections
(namely, part of NLO + NNLO + ... terms corresponding to real initial-state gluon
emissions) that can be taken into account in the form of transverse 
momentum dependent (TMD, or unintegrated) parton distributions\footnote{See reviews\cite{13,14} for more information.}.
It has become a widely exploited tool and it is of interest
and importance to test it in as many cases as possible. 

In the present note we apply the $k_T$-factorization approach to the 
associated $\gamma$ + jet production at LHC energies, which continues the line of our previous studies\cite{15,16,17}
where we have inspected inclusive photon as well as 
associated prompt photon (or rather Z boson) and heavy quark jet production.
Note that the associated $\gamma$ + jet production was already examined
in the $k_T$-factorization framework\cite{18,19}.
In particular, some photon-jet correlations have been studied at the RHIC and Tevatron energies\cite{18}.
However, initial state parton showers, which are important for the proper jet 
determination in the $k_T$-factorization approach, have been not 
taken into account in those calculations.
A simple model\cite{20} to implement the effects of parton showers 
into analytical calculations results in some difficulties in simultaneous 
description of photon transverse momentum and rapidity distributions in the whole kinematical range\cite{19}.
The importance of parton shower contributions to jet production was pointed out\cite{21}
and the method\cite{22} to reconstruct correctly the kinematics of the jets with taking into account TMD 
parton showers was proposed.
The major goal of the present article is to apply the method\cite{22} 
to associated $\gamma$ + jet production and improve our previous results\cite{19}
using a TMD shower implemented in the Monte-Carlo event generator \textsc{cascade}\cite{23}.  
Our other goal is the selection of TMD gluon densities in a proton
best suited to describe the available experimental data.

The outline of the paper is the following. In Section~2 we briefly
describe our approach. In Section~3 we
present the results of our calculations and confront them with
the available data. Our conclusions are summarised in Section~4.

\section{Theoretical framework} \indent

Let us briefly describe the calculation steps. We start from the off-shell gluon fusion subprocess:
\begin{equation}
\label{gg}
g^*(k_1)+g^*(k_2)\to\gamma(p)+q(p_1)+\bar q(p_2),
\end{equation}

\noindent
where the momenta of all particles are given in the parentheses.
The corresponding gauge-invariant off-shell production amplitude was calculated earlier\cite{24,25} and 
implemented into the Monte-Carlo event generator \textsc{cascade}\cite{23}
and newly developed parton-level Monte-Carlo event generator \textsc{pegasus}\cite{26}.
All the details of these calculations were explained in\cite{24,25}. We only
mention here that evaluation of the off-shell matrix element involves a special gluon 
polarization sum rule:
\begin{equation}
\label{sumgluon}
\sum \epsilon^\mu\epsilon^{*\nu}=\frac{k^\mu_T k^\nu_T}{\mathbf k^2_T},
\end{equation}

\noindent
where $\epsilon$ is the gluon polarization vector and $k_T$ its non-zero transverse momentum. 
In the collinear limit ${\mathbf k}_T^2 \to 0$ this expression converges
to the ordinary one after averaging over the azimuthal angle.
In all other respects the calculations follow the standard QCD Feynman rules.

Following\cite{17}, in addition to off-shell gluon-gluon fusion we take into account several subprocesses
involving quarks in the initial state, namely:
\begin{gather}
\label{qvg}
q(k_1) + g(k_2)\to \gamma(p) + q(p_1),\\
\label{qbarq}
q(k_1) + \bar q(k_2)\to \gamma(p) + g(p_1),\\
\label{qQ}
q(k_1) + q^\prime(k_2)\to \gamma(p) + q(p_1) + q^\prime(p_2),\\
\label{qq}
q(k_1) + \bar q(k_2)\to \gamma(p) + q^\prime(p_1) + \bar q^\prime(p_2),
\end{gather}

\noindent
where the momenta of all particles are given in the parentheses.
Despite of the fact, that quark densities are typically much lower than the gluon density at LHC conditions, 
these processes may become important at very large transverse momenta (or, respectively, at large parton longitudinal 
momentum fraction $x$, which is needed to produce large $p_T$ events) where the quarks are less suppressed or can even 
dominate over the gluon density. Here we find it reasonable to rely upon collinear 
Dokshitzer-Gribov-Lipatov-Altarelli-Parisi (DGLAP) factorization 
scheme\cite{27}, which provides better theoretical grounds in the large-$x$ region.
So, we consider a combination of two techniques
with each of them being used at the kinematic conditions where it is best suitable
(gluon induced subprocess (\ref{gg}) at small $x$ and quark-induced subprocesses (\ref{qvg}) --- (\ref{qq}) at large
$x$ values). Such combined scheme was successfully applied to describe the associated production
of prompt photons (or $Z$ bosons) and heavy quark jets at the LHC\cite{16,17}.
The calculation of production amplitudes (\ref{qvg}) --- (\ref{qq}) is a very straightforward
and a cross-check of our results has been done using \textsc{madgraph} tool\cite{28}.

Note that numerically we keep only valence quarks in (\ref{qvg}) to avoid any double counting.
The calculations based on another TMD scenario, the Parton Branching (PB) approach\cite{29,30},
should include both sea and valence quark contributions.
Similarly to conventional DGLAP-based evaluations, in order to describe the data, 
the PB calculations have to involve 
a number of additional higher-order subprocesses
(such as $qg \to \gamma qg$ subprocess)
properly matched with leading order terms (see, for example,\cite{31}).
The relation between the PB and the CCFM scheme applied here (see below) is out of 
our present consideration.

It is well-known that photons may also originate from the so-called fragmentation processes of partons 
produced in the hard interaction. However, an isolation requirement which is applied in 
the measurements\cite{1,2,3,4,5,6}, significantly reduces the rate for these processes: 
after applying the isolation cuts such contributions amount only to about 10\% of the 
visible cross section. Therefore, below we will neglect the contributions from the fragmentation 
mechanisms\footnote{The isolation requirement and additional conditions which preserve our calculations from 
divergences have been specially discussed in\cite{24}.}. 

As usual, to calculate the contributions of quark-induced subprocesses (\ref{qvg}) --- (\ref{qq}) one
has to convolute the corresponding partonic cross sections $d\hat \sigma_{ab}$
with the conventional parton distribution functions $f_a(x,\mu^2)$ in a proton:
\begin{equation}
  \sigma = \int dx_1 dx_2 \, d\hat\sigma_{ab}(x_1,x_2,\mu^2)f_a(x_1,\mu^2)f_b(x_2,\mu^2),
\end{equation}

\noindent
where indices $a$ and $b$ denote quark and/or gluon and $x_1$ and $x_2$
are the longitudinal momentum fractions of the colliding protons.
In the case of off-shell gluon-gluon fusion (\ref{gg}) we employ the $k_T$-factorization formula:
\begin{equation}
\label{sigma_kt}
\sigma=\int dx_1 dx_2 d\mathbf k_{1T}^2 d\mathbf k_{2T}^2 d\hat\sigma_{gg}^*(x_1,x_2,\mathbf k_{1T}^2,\mathbf k_{2T}^2,\mu^2)f_g(x_1,\mathbf k_{1T}^2,\mu^2)f_g(x_2,\mathbf k_{2T}^2,\mu^2),
\end{equation}

\noindent
where $f_g(x,\mathbf k_{T}^2,\mu^2)$ is the TMD gluon density in a proton. 
A comprehensive collection of the latter can be found in the \textsc{tmdlib} package\cite{32}, which 
is a C++ library providing a framework and an interface to the different parametrizations.
In the present paper we have tested two latest sets (namely, JH'2013 set 1 and JH'2013 set 2) which 
were obtained\cite{33} from the numerical solution of the CCFM gluon evolution equation. 
The CCFM equation provides a suitable tool since it smoothly interpolates between the
small-$x$ BFKL gluon dynamics and high-$x$ DGLAP one. 
The input parameters of the initial gluon distribution were fitted from the
best description of the precision DIS data on the proton structure functions $F_2(x,Q^2)$ and $F_2^c(x,Q^2)$.
For the conventional quark and gluon densities we used the MSTW’2008 (LO) set\cite{34}.
Numerical calculations at the parton level in the $k_T$-factorization approach and  
collinear QCD factorization were performed using the Monte-Carlo event generator \textsc{pegasus}.

A last important point of our calculations is connected with the proper determination of associated jet four-momentum:
the quarks and gluons produced in the hard subprocesses (1), (3) --- (6) can form
final-state hadronic jets. In addition to that, the produced photon is accompanied by a number
of gluons radiated in the course of the non-collinear evolution, which also give
rise to final jets. From all of these hadronic jets we choose the one (i.e. leading jet), carrying the largest
transverse momentum (and satisfying the experimental cuts) and then compute the cross-section of $\gamma$ + jet production.
Technically, we produce a Les Houche Event file\cite{35} in our parton level calculations 
performed using the Monte-Carlo event generator \textsc{pegasus} and then process the file with a 
TMD shower implemented in \textsc{cascade}, 
thus fully reconstructing the CCFM evolution.
This approach gives us the possibility to take into account 
the contributions from initial state parton showers in a consistent way
and, of course, essentially differs from simple model\cite{20} used in the previous calculations\cite{19}.
This model\cite{20} was based on the assumption that the gluon, emitted in the 
last non-collinear evolution step, compensates the whole transverse momentum of the gluon participating 
in the hard subprocess. Under this assumption, all the other emitted gluons can be collected together
in the proton remnant, which carries only a negligible transverse momentum (see\cite{20} for more information).
Concerning the quark-induced subprocesses (\ref{qvg}) --- (\ref{qq}), calculated in the 
conventional (collinear) QCD factorization, we used the latest version of \textsc{pythia} package\cite{36} to 
process the Les Houche Event files generated by \textsc{pegasus}.
The jets are reconstructed with the anti-$k_T$ algorithm, 
implemented in the \textsc{FastJet} tool\cite{37}. 

\section{Numerical results} \indent

Throughout this paper, all calculations are based on the following parameter setting.
In collinear QCD factorization we use one-loop strong coupling with
$n_f = 4$ massless quark flavors and $\Lambda_{\rm QCD} = 200$~MeV, the factorization and renormalization 
scales are both set equal to the produced photon transverse energy, $\mu_R = \mu_F = E_T^{\gamma}$.
In the $k_T$-factorization calculations we use a two-loop expression for the strong coupling
(as it was originally done in the fit\cite{33}), set $\mu_R = E_T^{\gamma}$ and define the factorization scale as
$\mu_F^2 = \hat s + {\mathbf Q}_T^2$ with $\hat s$ and ${\mathbf Q}_T$
being the subprocess invariant energy and the net transverse momentum of the initial 
off-shell gluon pair, respectively.
Note that the definition of $\mu_F$ is dictated by the
CCFM evolution algorithm\cite{33}.

The measurements of associated $\gamma$ + jet production cross sections have been carried out by 
the CMS\cite{1,2} and ATLAS\cite{3,4,5,6} Collaborations at LHC energies $\sqrt s = 7$, $8$ and $13$~TeV.
However, the data\cite{2,5,6} refer to the region of high $E_T^\gamma$ (i.e., region of relatively large $x \sim E^\gamma_T/\sqrt s$),
where standard quark-induced subprocesses (\ref{qvg}) --- (\ref{qq}) dominate.
We do not analyse events of this kind in the present study and only concentrate on the small 
and moderate $E_T^\gamma$ data\cite{1,3,4},
where off-shell gluon-gluon fusion plays a role (see discussion below).
The experimental acceptance, anti-$k_T$ algorithm 
radius $R^{\rm jet}$ and $\eta -\phi$ separation $\Delta R^{\gamma - \rm jet}$ 
implemented in the experimental analyses\cite{1,3,4} are collected in Table~1.
The CMS Collaboration has reported\cite{1} the measurements of triple-differential cross section 
$d\sigma/dE^\gamma_T d\eta^\gamma d\eta^{\rm jet}$ for various configurations of the photon and leading jet
at $\sqrt s = 7$~TeV. In the ATLAS analysis\cite{3}, the differential cross section
$d\sigma/dE_T^\gamma$ has been measured for three different rapidity ranges of the leading jet:
$|y^{\rm jet}| < 1.2$, $1.2 < |y^{\rm jet}| < 2.8$ and $2.8 < |y^{\rm jet}| < 4.4$.
For each rapidity configuration the same-sign ($\eta^\gamma \eta^{\rm jet} > 0$)
and opposite-sign ($\eta^\gamma \eta^{\rm jet} < 0$) cases are
studied separately.
More recently, the ATLAS Collaboration has presented measurements\cite{4} of $\gamma$ + jet 
cross sections as a function of the photon transverse energy $E_T^\gamma$, leading jet 
transverse momentum $p_T^{\rm jet}$ and rapidity $y^{\rm jet}$ at the same energy $\sqrt s$. In addition,
the cross sections as a function of the difference between the azimuthal angles of the
photon and jet $\Delta \phi^{\gamma - \rm jet}$, invariant mass $m^{\gamma - \rm jet}$ and  
scattering angle $\cos \theta = \tanh (y^\gamma - y^{\rm jet})/2$ have been reported.

We confront our predictions with the CMS\cite{1} and ATLAS\cite{3,4} data in Figs.~1 --- 4. 
The results obtained using the JH'2013 set 1 and set 2 gluon densities (including the effects of both 
initial and final state parton showers) are plotted with 
scale uncertainties depicted as green and yellow shaded bands, respectively.
To estimate these uncertainties we used the JH'2013 set 1(2)$+$
and JH'2013 set 1(2)$-$ gluon distributions instead of default JH'2013 set 1(2) density. 
These two sets represent a variation of the renormalization scale used in
the off-shell production amplitude. The JH'2013 set 1(2)$+$ stands for a variation 
of $2\mu_R$, while JH'2013 set 1(2)$-$ refects $\mu_R/2$ (see\cite{33}).
To estimate the scale uncertainties in the quark-involving subprocesses (\ref{qvg}) --- (\ref{qq}), 
calculated in the collinear QCD factorization, we have varied the 
scales $\mu_R$ and $\mu_F$ by a factor of $2$ around their default values.
Separately we show the contribution of off-shell gluon-gluon fusion subprocess (\ref{gg}), 
calculated with $k_T$-factorization.

As one can see, we achieved good agreement of our predictions
with the CMS\cite{1} and ATLAS\cite{3,4} data in the whole kinematical region 
within the experimental and theoretical uncertainties.
The predictions from JH'2013 set 2 gluon are somewhat lower than those from JH'2013 set 1,
especially for the distribution in scattering angle $\theta$ (see Fig.~4).
The reason for this lies in the additional limitation of the phase space in these measurements\cite{4},
namely, $\cos\theta < 0.83$, $m^{\gamma - \rm jet} > 161$~GeV and $|y^\gamma+y^{\rm jet}| < 2.37$,
which moves the probed kinematical region to somewhat larger $x$. 
Note that the measured distribution in $\cos \theta$ is sensitive to the $\gamma$ + jet production dynamics
and well reproduced in our calculations with JH'2013 set 1 gluon.
One can see also that the off-shell gluon-gluon fusion subprocess (\ref{gg}), in which we are mainly interested, 
dominates at low and moderate transverse energy ($E_T^\gamma \leq 120$ or $150$~GeV) and practically 
does not contribute at larger values\footnote{We have tested also newly proposed TMD gluon 
densities\cite{38} obtained as solutions of DGLAP equations with keeping exact kinematics using PB 
method\cite{29,30}. The PB-based predictions (not shown in the Figs.~1 --- 4) significantly underestimate the data, that
demonstrates the role of small-$x$ resummation in the CCFM equation. Necessity of taking into account of 
higher-order QCD corrections when using the PB parton densities has been demostrated recently in\cite{39}.}.
So, the subleading quark-induced subprocesses (\ref{qvg}) --- (\ref{qq}) are
important to achieve an adequate description of the data in the whole $E_T^\gamma$ region.
Similar conclusions were made earlier in\cite{17} in the case of associated $Z$ boson
and heavy quark jet production at the LHC.

\begin{table}
\label{table1}
\begin{center}
\begin{tabular}{|c|c|c|c|}
\hline
 & CMS \cite{1}  & ATLAS \cite{3}  & ATLAS \cite{4} \\\hline
$E_T^\gamma$/GeV & $>40$ & $>45$ & $>25$  \\\hline
$p_T^{\text{jet}}$/GeV & $>30$ & $>40$ & $>20$  \\\hline
$|\eta^\gamma|$ & $<2.5$ & $<2.37$ (excl. $1.37<|\eta^\gamma|<1.52$)  & $<1.37$  \\\hline
$|\eta^{\text{jet}}|$ & $<2.5$ & $<2.37$ & $<4.4$  \\\hline
$\Delta R^{\gamma - \rm jet}$ & 0.5 & 1.0 & 1.0  \\\hline
$R^{\rm jet}$ & 0.5 & 0.6 & 0.4 \\\hline
\end{tabular}
\end{center}
\caption{The kinematical cuts and anti-$k_T$ algorithm 
radius $R^{\rm jet}$ implemented in the experimental analyses\cite{1,3,4} and in our calculations.}
\end{table}

As was noted above, the initial state parton shower in \textsc{cascade} is based on the CCFM
evolution equation, while the final state parton shower is based on the DGLAP equations.
To investigate the influence of parton showers in a final state for description
of the LHC data, we repeated the calculations with taking into account parton showers in initial state only.
These results are presented in Figs.~1 --- 4 by dashed histograms.
We find that the final state radiation effects are quite negligible in most of the distributions, 
excluding only the region of very small $\Delta\phi^{\gamma - \rm jet}$ (see Fig.~3).

As a last point of our study, we present results of our calculations 
where the simple model\cite{20} has been applied in the jet selection procedure, similar to previous
evaluations\cite{19} (dash-dotted histograms in Figs.~1 --- 4).
As one can see, the achieved overall description of the considered experimental data is systematically
worse, both in normalization and shape. Although the simple approach\cite{20} is able to 
describe more or less adequately the measured $E_T^\gamma$ distributions 
in the some kinematical region (as it is shown in Fig.~2),
it fails for more exclusive observables, such as $\Delta\phi^{\gamma - \rm jet}$ variable (see Fig.~3).
Thus, it indicates again the importance of 
taking into account contributions from initial state parton showers for the
proper determination of the leading jet in the $k_T$-factorization approach. 

\section{Conclusion} \indent

We have considered associated production of prompt photon and hadronic jets at LHC conditions. 
The calculations were performed in a ``combined'' scheme employing both
$k_T$-factorization and collinear factorization in QCD, with each of them used in the 
kinematic conditions of its best reliability. 
The dominant contribution is represented by the off-shell gluon-gluon fusion subprocess
$g^*g^* \to \gamma q\bar q$. Several subleading quark-induced subprocesses contributing at 
$\mathcal O(\alpha\alpha_s)$ and $\mathcal O(\alpha\alpha_s^2)$
have been taken into account in the conventional collinear scheme.
To reconstruct correctly the kinematics of the hadronic jets the TMD parton shower 
generator \textsc{cascade} has been applied.

Using the TMD gluon densities derived from the CCFM evolution equation, we have
achieved reasonably good agreement between our theoretical predictions and the 
CMS and ATLAS experimental data. 
We have demonstrated the importance of initial state parton showers
for jet determination in the $k_T$-factorization approach.

\section*{Acknowledgements} \indent

The authors thank S.P.~Baranov for very useful discussions and important remarks.
A.V.L. and M.A.M. are grateful to DESY Directorate for the support in the framework of
Cooperation Agreement between MSU and DESY on phenomenology of the LHC processes
and TMD parton densities. 
M.A.M. was also supported by a grant of the foundation 
for the advancement of theoretical physics and mathematics "Basis" 17-14-455-1.

\newpage

\begin{figure}
\begin{center}
\includegraphics[width=5.5cm]{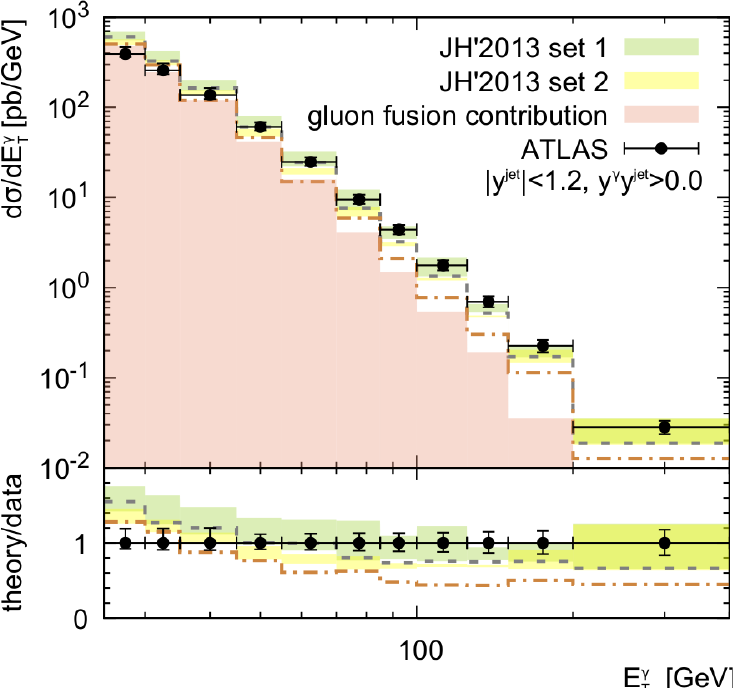} \hspace{1cm}
\includegraphics[width=5.5cm]{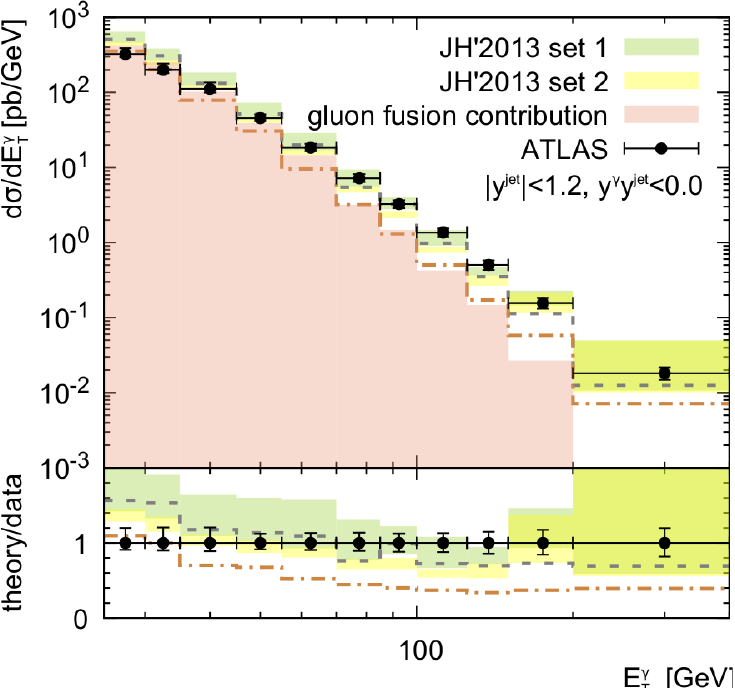}
\includegraphics[width=5.5cm]{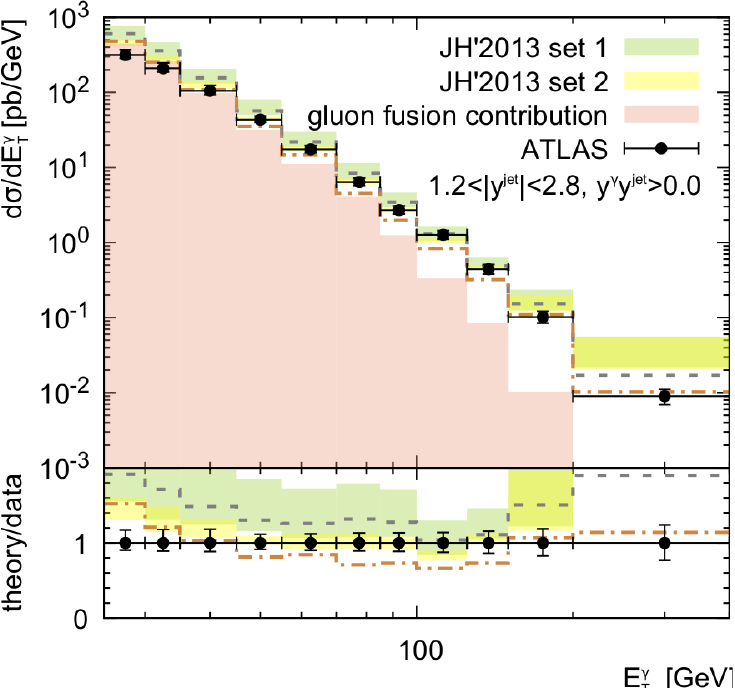} \hspace{1cm}
\includegraphics[width=5.5cm]{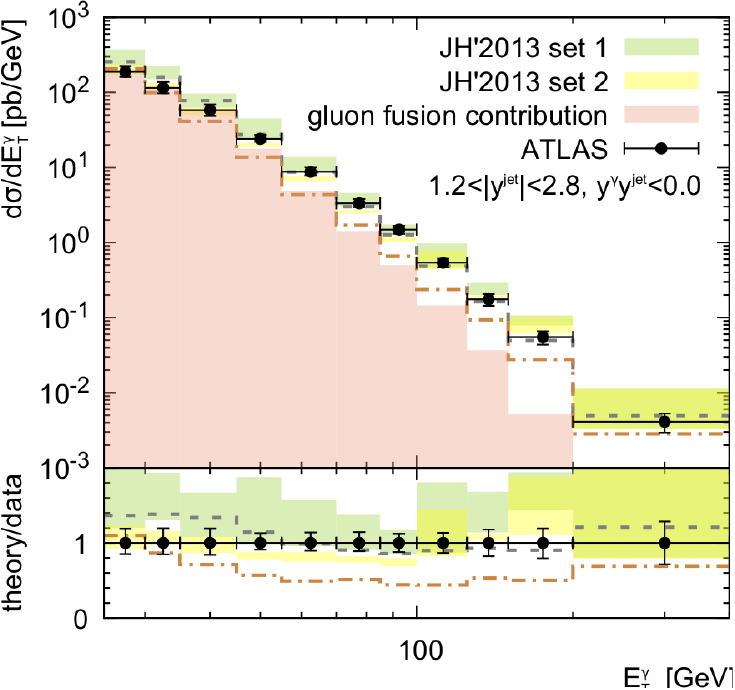}
\includegraphics[width=5.5cm]{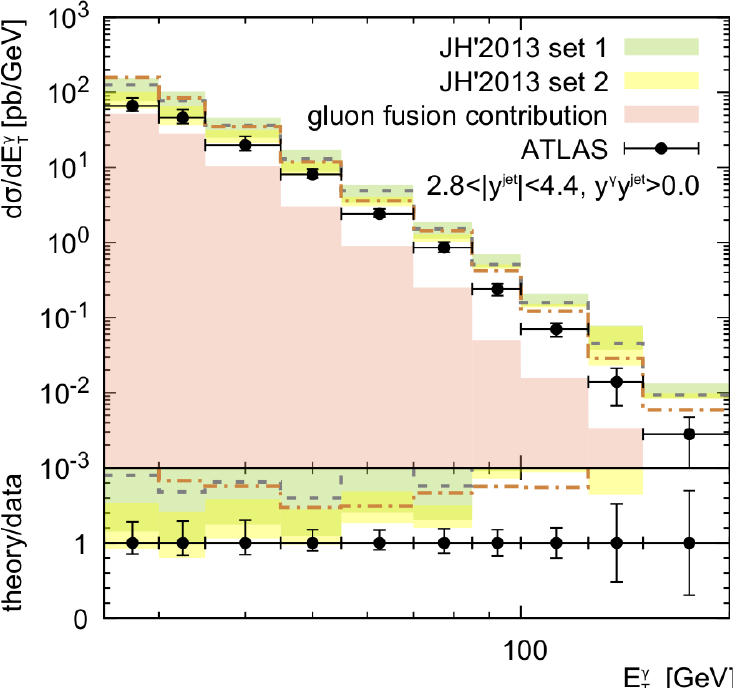} \hspace{1cm}
\includegraphics[width=5.5cm]{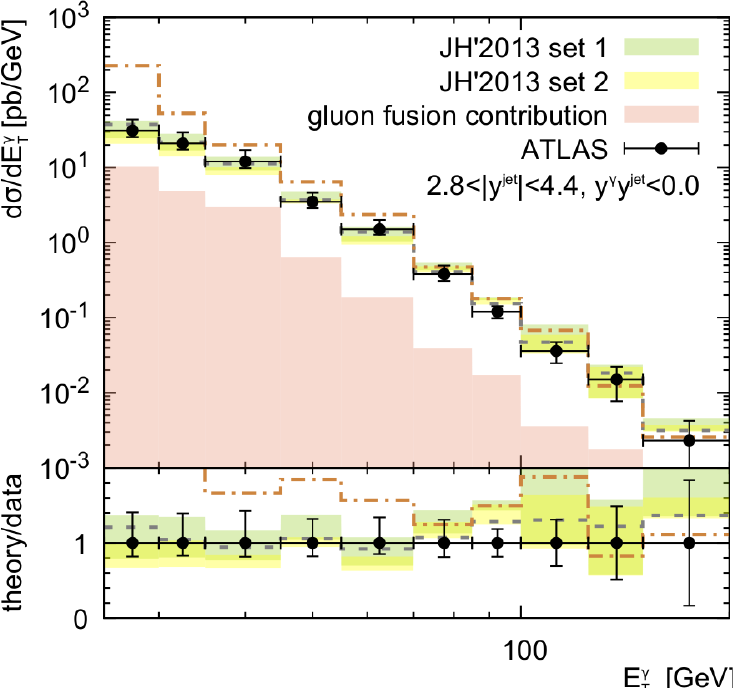}
\caption{The differential cross sections of associated $\gamma$ + jet production at $\sqrt s = 7$~TeV
as function of the prompt photon transverse energy $E_T^\gamma$ in different regions of rapidities.
The green and yellow shaded band represent the results obtained with JH'2013 set 1 and set 2 
gluon densities (with scale uncertainties). Dashed histograms corresponds to the 
predictions without final-state parton showers, dash-dotted histograms correspond to 
the results, obtained with simple approach\cite{20}.
Separately shown contribution from the off-shell gluon-gluon fusion subprocess~(1).
Everywhere the JH'2013 set 1 gluon density was used.
The experimental data are from ATLAS\cite{3}.}. 
\label{fig1}
\end{center}
\end{figure}

\begin{figure}
\begin{center}
\includegraphics[width=5.5cm]{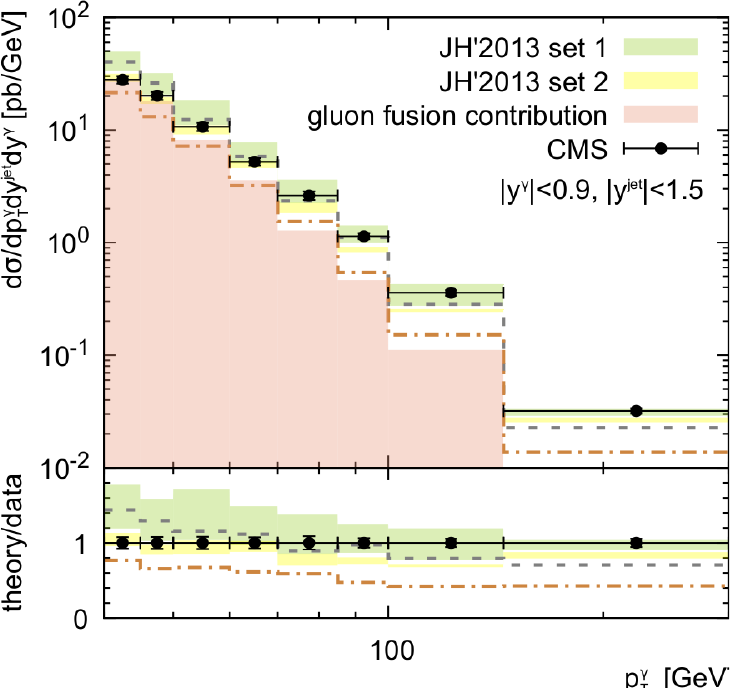} \hspace{1cm}
\includegraphics[width=5.5cm]{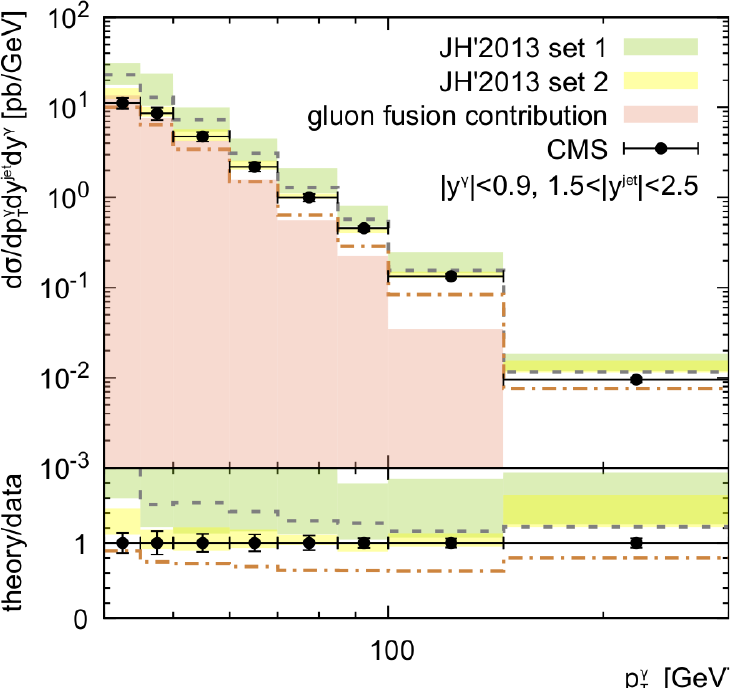}
\includegraphics[width=5.5cm]{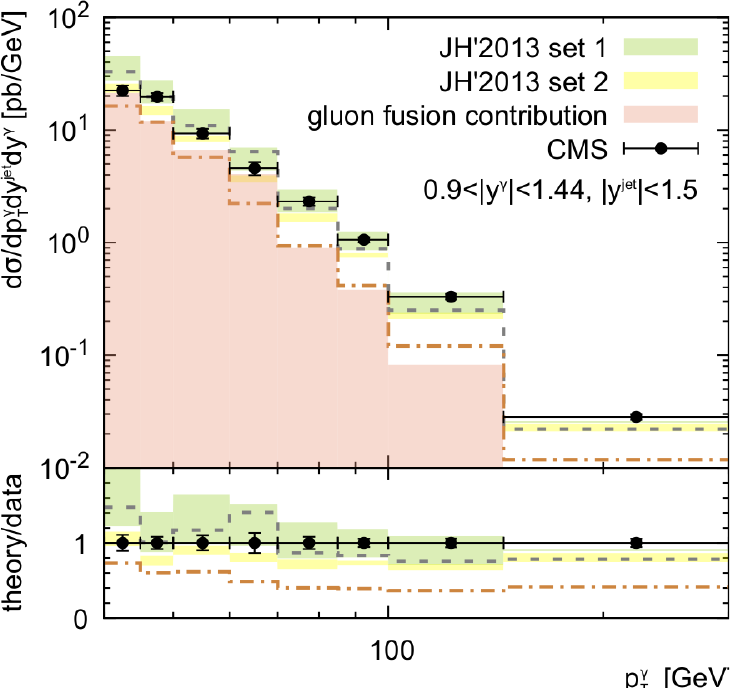} \hspace{1cm}
\includegraphics[width=5.5cm]{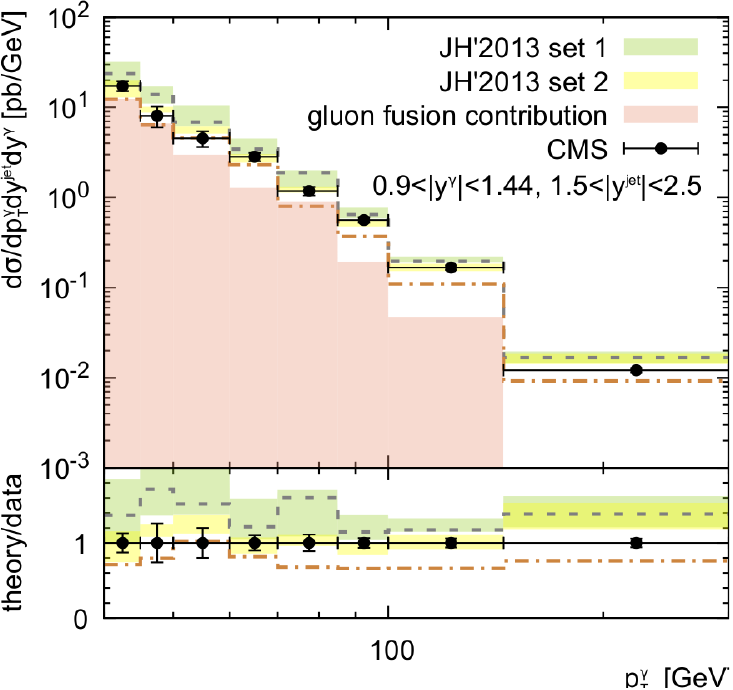}
\includegraphics[width=5.5cm]{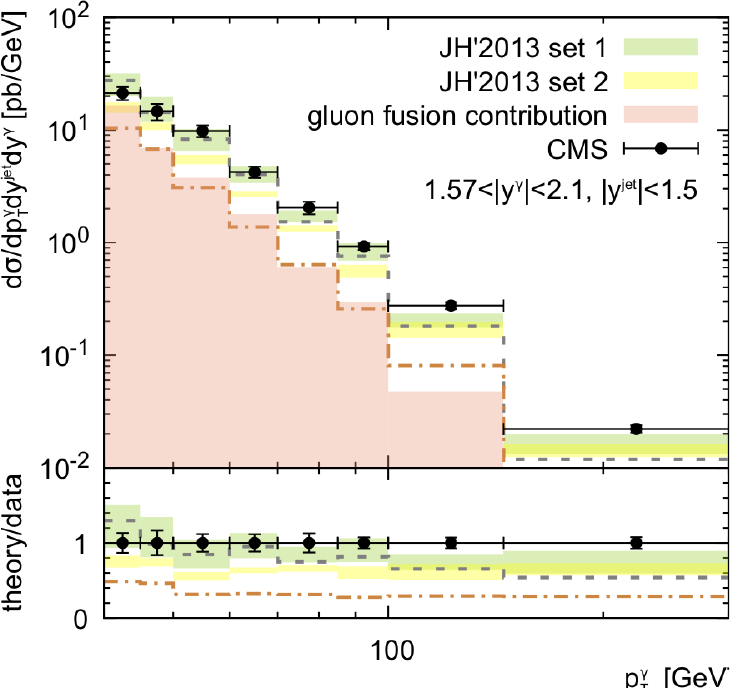} \hspace{1cm}
\includegraphics[width=5.5cm]{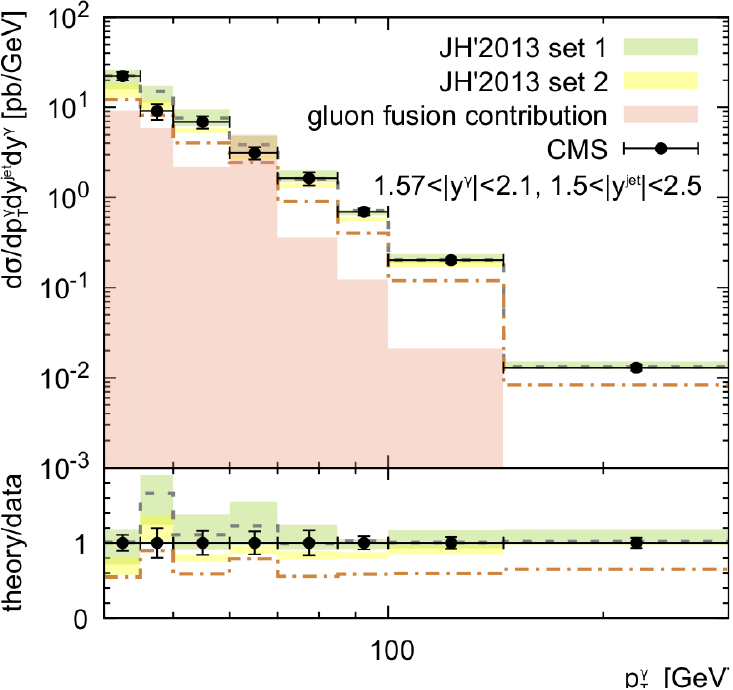}
\includegraphics[width=5.5cm]{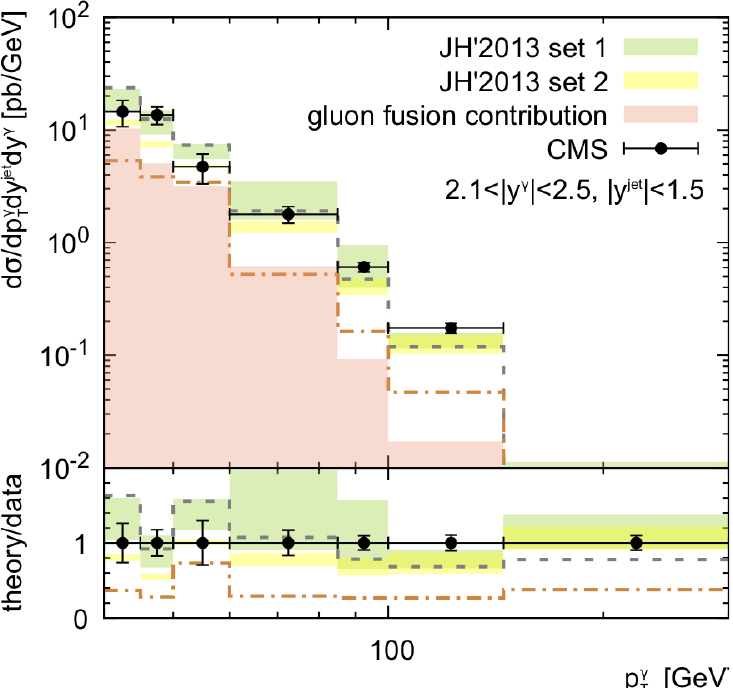} \hspace{1cm}
\includegraphics[width=5.5cm]{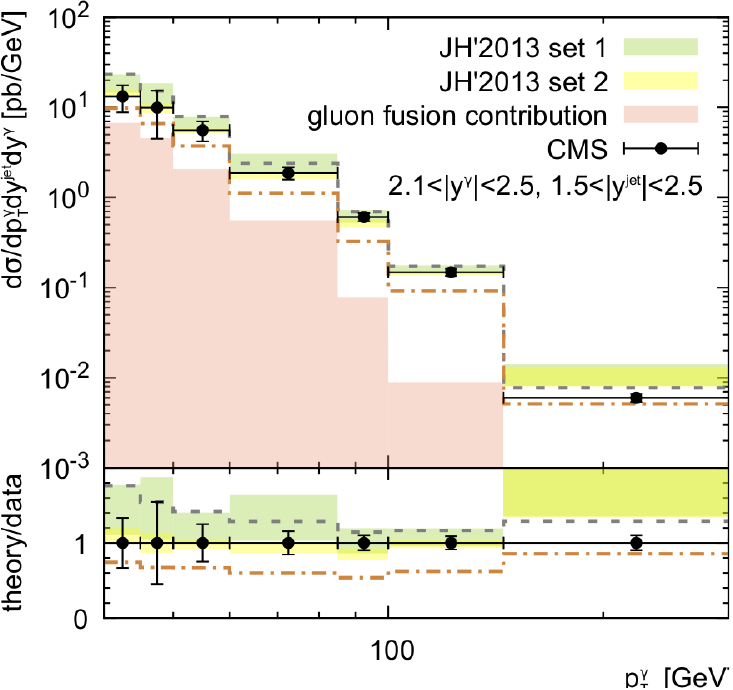}
\caption{The triple-differential cross sections of associated $\gamma$ + jet production at $\sqrt s = 7$~TeV
as function of the photon transverse energy in different regions of rapidities.
Notation of histograms is the same as in Fig.~1. 
The experimental data are from CMS\cite{1}.}
\label{fig2}
\end{center}
\end{figure}

\begin{figure}
\begin{center}
\includegraphics[width=5.5cm]{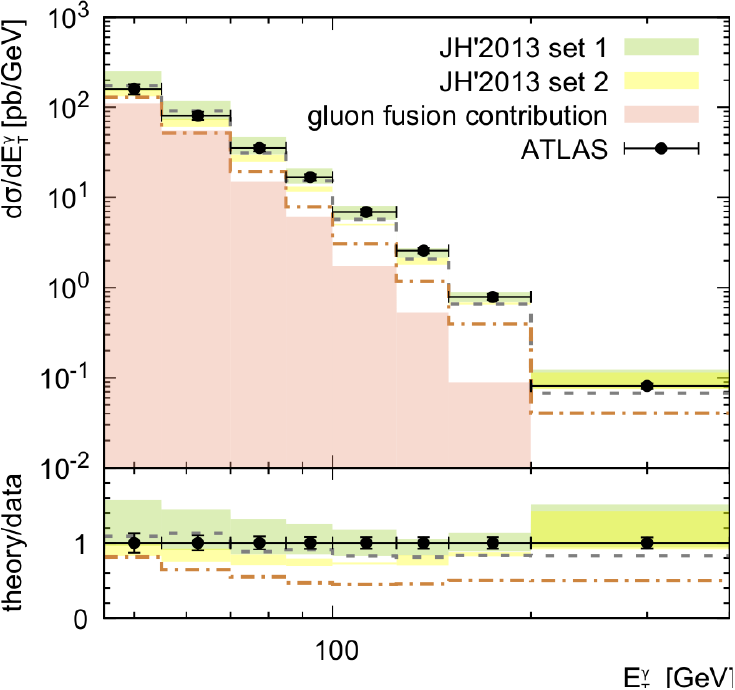} \hspace{1cm}
\includegraphics[width=5.5cm]{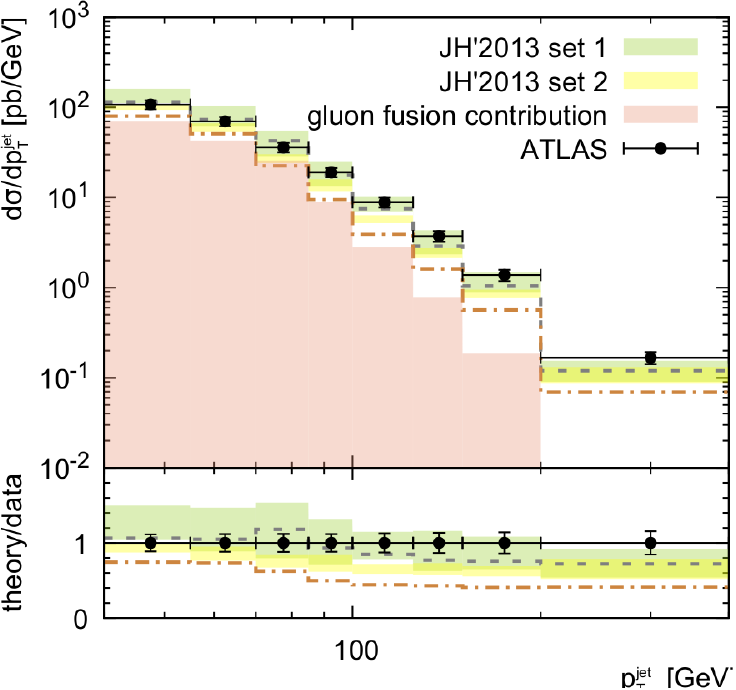}
\includegraphics[width=5.5cm]{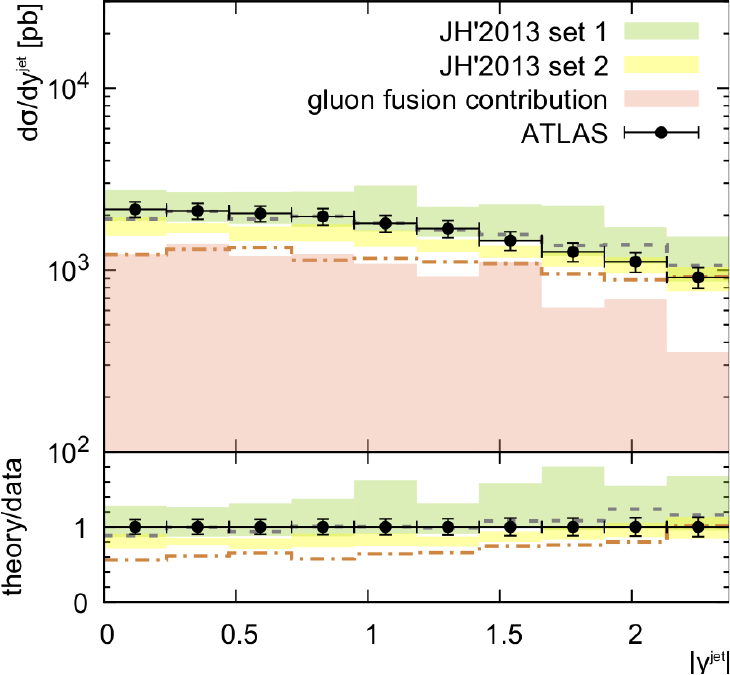} \hspace{1cm}
\includegraphics[width=5.5cm]{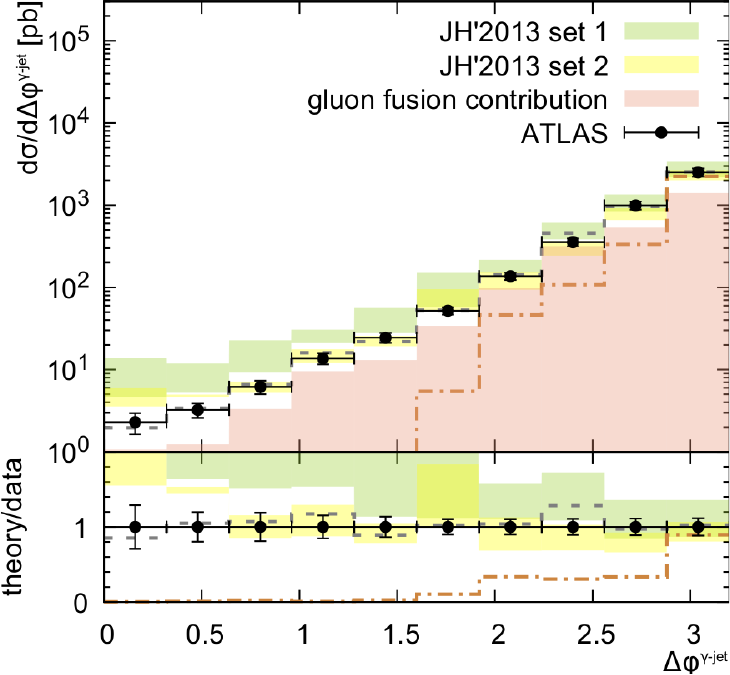}
\caption{The differential cross sections of associated prompt photon and jet production at $\sqrt s = 7$~TeV
as functions of photon transverse energy $E_T^\gamma$, jet transverse momentum $p_T^{\text {jet}}$, 
jet rapidity $y^{\text {jet}}$ and azimuthal angle difference between the prompt photon and the 
leading jet $\Delta\phi$. Notation of histograms is the same as in Fig.~1. 
The experimental data are from ATLAS\cite{4}.}
\label{fig3}
\end{center}
\end{figure}

\begin{figure}
\begin{center}
\includegraphics[width=5.5cm]{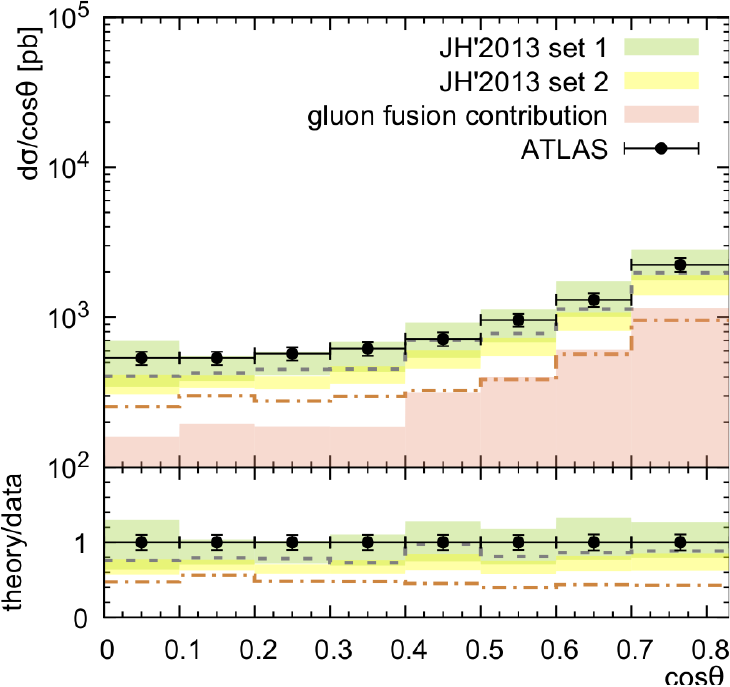} \hspace{1cm}
\includegraphics[width=5.5cm]{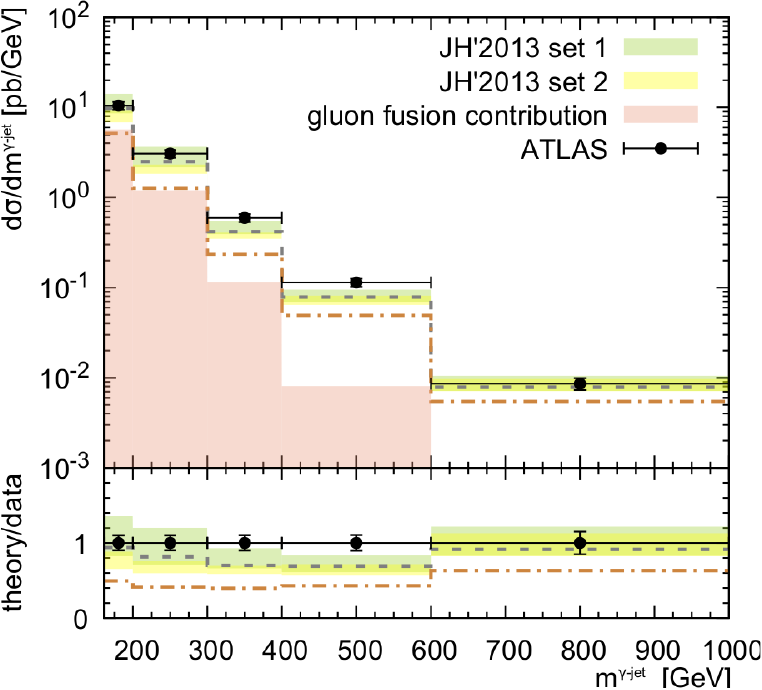}
\caption{The differential cross sections of associated prompt photon and jet production at $\sqrt s = 7$~TeV
as function of scattering angle $\cos\theta$ and the invariant mass of the prompt photon 
and the leading jet. Additional cuts $\cos\theta < 0.83$, $m^{\gamma - \rm jet} > 161$~GeV 
and $|y^\gamma+y^{\text {jet}}| < 2.37$ are applied.
Notation of histograms is the same as in Fig.~1. The experimental data are from ATLAS\cite{4}.}
\label{fig4}
\end{center}
\end{figure}

\end{document}